\documentclass[a4paper]{report}
\usepackage[utf8]{inputenc}
\usepackage[T1]{fontenc}
\usepackage{RJournal}
\usepackage{amsmath,amssymb,array}
\usepackage{booktabs}

\begin{document}

\sectionhead{Contributed research article}
\volume{XX}
\volnumber{YY}
\year{20ZZ}
\month{AAAA}

\begin{article}
\title{singR: An R package for Simultaneous non-Gaussian Component
Analysis for data integration}
\author{by Liangkang Wang, Irina Gaynanova, and Benjamin Risk}

\maketitle

\abstract{%
This paper introduces an R package that implements Simultaneous
non-Gaussian Component Analysis for data integration. SING uses a
non-Gaussian measure of information to extract feature loadings and
scores (latent variables) that are shared across multiple datasets. We
describe and implement functions through two examples. The first example
is a toy example working with images. The second example is a simulated
study integrating functional connectivity estimates from a resting-state
functional magnetic resonance imaging dataset and task activation maps
from a working memory functional magnetic resonance imaging dataset. The
SING model can produce joint components that accurately reflect
information shared by multiple datasets, particularly for datasets with
non-Gaussian features such as neuroimaging.
}

\hypertarget{introduction}{%
\section{Introduction}\label{introduction}}

Combining information across different data sets collected on the same
participants can improve our understanding of biology and diseases. In
neuroimaging, combining information across different modalities, or
types of imaging data, can lead to a more comprehensive picture of human
brain activity. Commonly used modalities to investigate brain function
include functional magnetic resonance imaging (fMRI), resting-state fMRI
(rs-fMRI), diffusion MRI, structural images, electroencephalography, and
positron emission tomography. Combining data from multiple modalities
can result in a better understanding of the underlying biology than
analyzing each modality separately \citep{RN176}. In previous studies,
researchers use data fusion to define a set of joint components that are
composed of subject scores (a vector in \(\mathbb{R}^n\), where \(n\) is
the number of subjects) and loadings (a vector in \(\mathbb{R}^{p_k}\),
where \(p_k\) is the number of variables in the \(k\)th dataset). For a
particular component, the subject scores are equal or strongly
correlated across datasets. The loadings represent the relative
importance of a variable to each component. A higher subject score
implies the vector of loadings is more important in that individual.

Data integration approaches for neuroimaging should accommodate the
distinct statistical properties of imaging data. Imaging features
characterizing brain activation and intrinsic functional connectivity
are substantially more non-Gaussian compared to noise. Methods employing
principal component analysis (PCA) for dimension reduction and
independent component analysis (ICA) have been widely used in
neuroscience \citep{RN176, RN179, 7373530}. ICA maximizes the
non-Gaussianity of components, which is useful for extracting
interesting features from imaging data. ICA is commonly used to estimate
resting-state networks \citep{beckmann2005investigations}, estimate
task-activated components \citep{SUI2010123}, and has been used to
derive network structure in resting-state correlation matrices
\citep{AMICO2017201}. \citet{RN146} proposed simultaneous non-Gaussian
component anaylsis (SING) for analyzing two datasets, which uses an
objective function that maximizes the skewness and kurtosis of latent
components with a penalty to enhance the similarity between subject
scores. Unlike previous methods, SING does not use PCA for dimension
reduction, but rather uses non-Gaussianity, which can improve feature
extraction.

Some useful software have been developed in this area. In multimodal
analysis and multitask fMRI data fusion, \pkg{FIT} is a Matlab toolbox
that implements jointICA, parallel ICA \citep{RN184}, and
multimodal/multiset CCA \citep{RN183}. \pkg{GIFT} provides functions for
conducting group ICA on fMRI data from multiple subjects from a single
modality \citep{RN180, RN181, RN182}. On the Comprehensive R Archive
Network (CRAN), there are several R packages for ICA functions,
including \CRANpkg{steadyICA} \citep{steadyICA}, \CRANpkg{ica}
\citep{ica}, \CRANpkg{fastICA} \citep{fastICA}, \CRANpkg{JADE}
\citep{JADE}, and \CRANpkg{templateICAr} \citep{templateICAr}. These R
packages use different algorithms to extract non-Gaussian features but
are designed for decomposing a single modality. For data integration,
\textbf{r.jive} \citep{r.jive} and \textbf{ajive}
\citep{ajivepackage, feng2018angle} capture the joint variation, or
variance shared between datasets, and individual variation, or variance
unique to a dataset. JIVE methods use singular value decompositions,
which are related to maximizing variance instead of non-Gaussianity. In
this way, there exists a need for freely available software for
extracting joint structure from multiple datasets using non-Gaussian
measures of information.

This paper develops \CRANpkg{singR}, an R package to implement
\citet{RN146}. Similar to \textbf{r.jive}, the input data to SING are
\(n \times p_k\) data matrices, \(k=1,2\), where each row is a vector of
features from a subject. Unlike existing software, \CRANpkg{singR} uses
non-Gaussianity for both dimension reduction and feature extraction.

This paper is structured as follows. In Section {[}2{]}, we review the
Simultaneous non-Gaussian component analysis (SING) model. In Sections
{[}3{]} and {[}4{]}, we present the main functions in \CRANpkg{singR}
package and show how to utilize it for joint components estimation in
two example datasets. Finally, Section {[}5{]} summarizes our
conclusions.

\hypertarget{methods}{%
\section{Methods}\label{methods}}

\hypertarget{linear-non-gaussian-component-analysis}{%
\subsection{Linear non-Gaussian Component
Analysis}\label{linear-non-gaussian-component-analysis}}

\emph{Matrix decomposition for one dataset.} Based on linear
non-Gaussian component analysis (LNGCA), we first summarize a matrix
decomposition for a single dataset \(X\in \mathbb{R} ^{n\times p_{x}}\)
(\(n\) subjects and \(p_{x}\) features) into a non-Gaussian subspace and
a Gaussian subspace. Each row of \(X\) is a vector of features from the
\(i\)th subject. Let \(X_{c}\) denote the double-centered data matrix
such that \(1^{T}X_{c}=0^{T}\) and \(X_{c}1=0\) where \(\mathbf{1}\)
denotes the vector of ones of appropriate dimension, which has rank
\(n-1\) when \(p_{x}>n\). Let \(I_{r_{x}}\) denote the
\(r_{x}\times r_{x}\) identity matrix. Then define the matrix
decomposition

\begin{equation}
X_{c}=M_{x}S_{x}+M_{N_{x}}N_{x}
\label{eq:decomp}
\end{equation}

where:\\
1. \(M_{x}\in \mathbb{R}^{n\times r_{x}}\) and
\(M_{N_{x}}\in \mathbb{R}^{n\times (n-r_{x}-1)}\). The columns of
\(M_{x}\) are called subject scores, and the matrix
{[}\(M_{x}\),\(M_{N_{x}}\){]} is called the mixing matrix and has rank
\(n-1\).\\
2. \(S_{x} \in \mathbb{R}^{r \times p_{x}}\) and
\(N_{x} \in \mathbb{R}^{n-r_{x}-1 \times p_{x}}\).
\(S_{x}S_{x}^{T}=p_{x}I_{r_{x}}\),\(N_{x}S_{x}^{T}=0_{(n-r_{x}-1)\times r_{x}}\).
The rows of \(S_{x}\) are the non-Gaussian components, and elements of
\(S_{x}\) are called variable loadings because \(X_{c}S_{x}^{T}=M_{x}\).
The rows of \(N_{x}\) are the Gaussian components.\\
3. The rows of \(S_{x}\) have the largest non-Gaussianity, as described
below.

This decomposition may be meaningful in neuroimaging studies because: 1)
vectorized components like brain activation maps and resting-state
networks have highly non-Gaussian distributions, and 2) the
decomposition is useful when \(p_x \gg n\), e.g., the number subjects is
smaller than the number of voxels or edges.

To achieve the matrix decomposition in \eqref{eq:decomp}, we need to find
a separating matrix \(A_{x}\) that maximizes non-Gaussianity and
satisfies the constraint
\(A_{x}X_{c}X_{c}^{T}A_{x}^{T}=S_{x}S_{x}^{T}=p_{x}I_{r_{x}}\). We
utilize a prewhitening matrix and then reparameterize the model with a
semiorthogonal separating matrix to enforce this constraint. Let
\(\widehat{\Sigma}_{x}=X_{c} X_{c}^{\top} / p_{x}\). Then define the
eigenvalue decomposition
\(\widehat{\Sigma}_{x}=V_{x} \Lambda_{x} V_{x}^{\top}\) and the
prewhitening matrix
\(\widehat{L}_{x}=V_{x} \Lambda_{x}^{-1 / 2} V_{x}^{\top}\). (For
double-centered \(X_c\), this is understood to be the square root of the
generalized inverse from the non-zero eigenvalues.) Note
\(A_{x}=U_{x} \widehat{L}_{x}\), and it follows that
\(\hat{M}_x = \widehat{L}_x^{-}U_x^\top\), with \(\widehat{L}_x^-\)
denoting the generalized inverse. Let \(f()\) be a measure of
non-Gaussianity. Then \eqref{eq:decomp} is estimated using

\begin{equation}
\begin{split}
& \underset{U_{x}}{\text{minimize}}-\sum_{l=1}^{r_{x}} f\left(u_{x l}^{\top} \widehat{L}_{x} X_{c}\right),\\ 
& \text{subject to}\quad   U_{x} U_{x}^{\top}=I_{r_{x}},
\end{split}
\label{eq:minU}
\end{equation}

where \(u_{x l}^{\top}\) is the \(l\)th row of the \(r_{x}\times n\)
matrix \(U_{x}\).

We measure non-Gaussianity with the \textbf{Jarque-Bera (JB) statistic},
a combination of squared skewness and kurtosis. For a vector
\(s \in \mathbb{R} ^{p}\), the JB statistic is

\begin{equation}
f(s)=0.8\left(\frac{1}{p} \sum_{j} s_{j}^{3}\right)^{2}+0.2\left(\frac{1}{p} \sum_{j} s_{j}^{4}-3\right)^{2}.
\label{eq:jbstatistics}
\end{equation}

\hypertarget{simultaneous-non-gaussian-component-analysis-model}{%
\subsection{Simultaneous non-Gaussian component analysis
model}\label{simultaneous-non-gaussian-component-analysis-model}}

\emph{Matrix decomposition for two datasets.} We now decompose
\(X \in \mathbb{R} ^{n \times p_{x}}\) and
\(Y \in \mathbb{R} ^{n \times p_{y}}\) into a joint non-Gaussian
subspace defined by shared subject score directions, individual
non-Gaussian subspaces, and Gaussian subspaces. Let \(r_{j}\) denote the
rank of the joint non-Gaussian subspace and \(r_{x},\;r_{y}\) denote the
rank of the non-Gaussian subspaces for \(X\) and \(Y\), respectively.
Define the double-centered \(X_{c}\) and \(Y_{c}\), i.e.,
\(\mathbf{1}^\top X_c = 0\) and \(X_c \mathbf{1} = 0\). In data
applications, we also recommend standardizing each feature to have unit
variance, as common in PCA. The double centering with standardization
requires an iterative algorithm that standardizes each feature (mean 0
variance 1 across subjects), then centers the features for a given
subject (the mean of the features for a subject is 0), and repeats
(typically \(<10\) iterations on real data suffices). The function
\texttt{standard} is described in Section {[}3{]}.

The SING matrix decomposition is

\begin{equation}
\begin{split}
& X_{c}=M_{J}D_{x}S_{Jx}+M_{Ix}S_{Ix}+M_{Nx}N_{x},\\
& Y_{c}=M_{J}D_{y}S_{Jy}+M_{Iy}S_{Iy}+M_{Ny}N_{y},
\end{split}
\label{eq:two}
\end{equation}

where:\\
1. \(M_{J}\in \mathbb{R}^{n\times r_{J}}\),
\(M_{I_{x}}\in \mathbb{R}^{n\times (r_{x}-r_{J})}\),
\(M_{I_{y}}\in \mathbb{R}^{n\times (r_{y}-r_{J})}\),
\(M_{N_{x}}\in \mathbb{R}^{n\times (n-r_{x}-1)}\), and
\(M_{N_{y}}\in \mathbb{R}^{n\times (n-r_{y}-1)}\).\\
2. \(D_{x}\) and \(D_{y}\) are diagonal and allow the size of \(M_J\) to
vary between datasets.\\
3. \(S_{Jx}\) are the joint non-Gaussian components, \(S_{Ix}\) are the
individual non-Gaussian components, and \(N_{x}\) are the individual
Gaussian components, respectively, for \(X\), with constraints
\(S_{Jx}S_{Jx}^{T}=p_{x}I_{r_{J}}\),
\(S_{Ix}S_{Ix}^{T}=p_{x}I_{r_{x}-r_{J}}\),
\(S_{Jx}S_{Ix}^{T}=0_{r_J\times (r_{x}-r_J)}\),
\(N_{x}S_{Jx}^{T}=0_{(n-r_{x}-1)\times r_J}\),
\(N_{x}S_{Ix}^{T}=0_{(n-r_{x}-1)\times (r_{x}-r_J)}\), and similarly
define the components of \(Y\).

\emph{Simultaneous non-Gaussian Component Analysis fitting and
algorithm.} Recall the whitening matrix for \(X_{c}\) is
\(\widehat{L}_{x}\), and define its generalized inverse
\(\widehat{L}_{x}^{-}=(X_{c}X_{c}^{T}/p_{x})^{1/2}=V_{x} \Lambda_{x}^{1/ 2} V_{x}^{\top}\).
We will estimate a semiorthogonal unmixing matrix \(\widehat{U}_x\) such
that \(\widehat{M}_{x} = \widehat{L}_{x}^{-}\widehat{U}_{x}^{T}\).
Similarly define the whitening matrix \(\widehat{L}_{y}\) and
\(\widehat{M}_{y}\) for \(Y_{c}\). Let \(f\) be the JB statistic, as
defined in \eqref{eq:jbstatistics}. We consider

\begin{equation}
\begin{split}
& \underset{U_{x}, U_{y}}{\text{minimize}}-\sum_{l=1}^{r_{x}} f(u_{x l}^{\top} \widehat{L}_{x}X_{c})-\sum_{l=1}^{r_{y}} f(u_{y l}^{\top} \widehat{L}_{y}Y_{c})+\rho \sum_{l=1}^{r_{J}} d(\widehat{L}_{x}^{-} u_{x l}, \widehat{L}_{y}^{-} u_{y l}) \\
& \text{subject to } \quad  U_{x} U_{x}^{\top}=I_{r_{x}},\; U_{y} U_{y}^{\top}=I_{r_{y}},
\end{split}
\label{eq:curvilinear}
\end{equation}

where \(d(x,y)\) is the chosen distance metric between vectors \(x\) and
\(y\), calculated using the chordal distance:
\(d(x, y)=\left\|\frac{x x^{\top}}{\|x\|_{2}^{2}}-\frac{y y^{\top}}{\|y\|_{2}^{2}}\right\|_{F}^{2}\).
When columns are mean zero, the chordal distance between joint scores in
the SING objective function is equal to zero when their correlation is
equal to one. Larger values of the tuning parameter \(\rho\) result in
common \(M_J\), but smaller values result in highly correlated joint
structure and could also be considered. In our examples, we match
components from the separate LNGCA, determine candidate joint components
from a permutation test, and then set \(\rho\) equal to the sum of the
JB statistics of all candidate joint loadings divided by 10, which
results in
\(\widehat{L}_{x}^{-} u_{x l} \approx \widehat{L}_{y}^{-} u_{y l}\),
i.e., a shared \(M_J\).

Let \(\widehat{U}_{x}\) and \(\widehat{U}_{y}\) be the estimated value
of \(U_{x}\) and \(U_{y}\) in (5). The corresponding estimated
non-Gaussian components are defined as
\(\widehat{S}_{x}=\widehat{U}_{x}X_{w}\) and
\(\widehat{S}_{y}=\widehat{U}_{x}Y_{w}\), where
\(X_{w}=\widehat{L}_{x}X_{c}\), and \(Y_{w}=\widehat{L}_{y}Y_{c}\),
respectively. Then the first \(r_{J}\) columns of
\(\widehat{M}_{x}=\widehat{L}_{x}^{-}\widehat{U}_{x}^{\top}\), scaled to
unit norm, define \(\widehat{M}_{Jx}\). We can similarly define
\(\widehat{M}_{Jy}\). For sufficiently large \(\rho\),
\(\widehat{M}_{Jx}=\widehat{M}_{Jy}=\widehat{M}_J\), and more generally,
\(\widehat{M}_J\) is defined from their average. Additionally, the first
\(r_{J}\) rows of \(\widehat{S}_{x}\) correspond to
\(\widehat{S}_{Jx}\).

\hypertarget{overview-of-r-package-singr}{%
\section{Overview of R package
singR}\label{overview-of-r-package-singr}}

The R package \CRANpkg{singR} implements simultaneous non-Gaussian
component analysis for data integration in neuroimaging. Highlighted
below are key functions:

\textbf{lngca}: This function estimates non-Gaussian components for a
single dataset. Non-Gaussian components can be estimated using the
Jarque-Bera test statistic, which is the non-Gaussian measure used in
SING, or using likelihood component analysis, which achieves dimension
reduction while estimating the densities of non-Gaussian components
\citep{risklngca}. It returns \(\widehat{U}_{x}\) and
\(\widehat{S}_{x}\) from decomposing \(X_{c}\) through (2). It also
returns the non-gaussianity of each estimated component.

\textbf{standard}: This function is an iterative algorithm that
standardizes each feature (mean 0 variance 1 across subjects), then
centers the features for a given subject (the mean of the features for a
subject is 0) for the original datasets (\(\mathbb{R}^{n\times p}\)),
and repeats until the variance is approximately equal to 1 (typically
\(<10\) iterations on real data suffices).

\textbf{est.M.ols}: This function returns \(\widehat{M}_{x}\) with input
\(\widehat{S}_{x}\) and \(X_{c}\) with the formula
\(\widehat{M}_{x}=X_{c}\widehat{S}_{x}^{\top}\).

\textbf{greedymatch}: This function reorders the columns in
\(\widehat{U}_{x}\) to match the columns (subject scores) in
\(\widehat{U}_{y}\) based on the chordal distances between corresponding
\(\widehat{M}_{x}\) and \(\widehat{M}_{y}\).

\textbf{permTestJointRank}: This function tests whether the correlation
between matched columns (subject scores) is significant and returns the
family wise error rate corrected p-values.

\textbf{\%\^{}\%}: Calculates the matrix exponential. For example,
\texttt{A\%\^{}\%0.5} returns a matrix square root. Used during
prewhitening.

\textbf{calculateJB}: This function calculates the sum of the JB
statistics across components and is useful for determining the size of
the penalty parameter \(\rho\) (sufficiently large \(\rho\) results in
the chordal distance between \(M_{Jx}\) and \(M_{Jy}\) equal to 0).
Assumes the variance of each row of the input S is equal to 1 and mean
of each row is equal to 0.

\textbf{curvilinear}: This function gives the final estimates of
\(\widehat{U}_{x}\) and \(\widehat{U}_{y}\) using the curvilinear
algorithm derived from (5). This is a pure R implementation but is slow.

\textbf{curvilinear\_c}: This implements the curvilinear algorithm in
C++, which is faster.

\textbf{NG\_number}: This is a wrapper function for \texttt{FOBIasymp}
from \CRANpkg{ICtest} \citep{ICtest} that can be used to estimate the
number of non-Gaussian components in a single dataset.

\textbf{signchange}: This function makes the skewness of each row of
\(\widehat{S}_{x}\) positive, which is useful for visualizing
non-Gaussian component loadings.

\textbf{singR}: This function integrates all the functions above. We can
use this function to estimate joint scores and loadings from two
datasets \(X\) and \(Y\) and optionally return the individual scores and
loadings.

\hypertarget{application}{%
\section{Application}\label{application}}

To illustrate the use of \CRANpkg{singR}, we provide two examples.

\hypertarget{example-1.-the-toy-datasets-decomposition}{%
\subsection{Example 1. The toy datasets
decomposition}\label{example-1.-the-toy-datasets-decomposition}}

The tutorial dataset \texttt{exampledata} are included in the
\CRANpkg{singR} package. We generate the SING model in \eqref{eq:two} as
follows. We generate joint subject scores
\(M_{J}=[m_{J1},m_{J2}]\in \mathbb{R}^{n\times2}\) with
\(m_{J1}\sim N(\mu_{1},I_{n}),m_{J2}\sim N(\mu_{2},I_{n})\),
\(\mu_{1}=(1_{24}^{\top},-1_{24}^{\top})^{\top}\) and
\(\mu_{2}=(-1_{24}^{\top},1_{24}^{\top})^{\top}\). We set \(D_{x}=I\)
and \(D_{y}=diag(-5,2)\) to have differences in both sign and scale
between the two datasets. We generate \(M_{Ix}\) and \(M_{Iy}\) similar
to \(M_{J}\) using iid unit variance Gaussian entries with means equal
to
\(\mu_{3y}=(-1_{6}^{\top},1_{6}^{\top},-1_{6}^{\top},1_{6}^{\top},-1_{6}^{\top},1_{6}^{\top}-1_{6}^{\top},-1_{6}^{\top})^{\top}\),
\(\mu_{4y}=(1_{24}^{\top},-1_{24}^{\top})^{\top}\),
\(\mu_{3x}=(-1_{12}^{\top},1_{12}^{\top},-1_{12}^{\top},1_{12}^{\top})^{\top}\),
\(\mu_{4x}=(1_{12}^{\top},-1_{12}^{\top},1_{12}^{\top},-1_{12}^{\top})^{\top}\).
These means result in various degrees of correlation between the columns
of the mixing matrices. For the Gaussian noise, we generate \(M_{Nx}\),
\(M_{Ny}\), \(N_{x}\) and \(N_{y}\) using iid standard Gaussian mean
zero entries.

Each row of \(S_{Jx}\) and \(S_{Ix}\) is a vectorized image. We can
reshape the loadings back to their image dimensions for visualization.
The loadings \(S_{Jx}\) are inspired by activation patterns found in
functional MRI, and similar simulations were considered in
\citep{risklngca}. The rows of \(S_{Jy}\) and \(S_{Iy}\) are formed from
the lower diagonal of a symmetric matrix, which are inspired by ICA of
correlation matrices \citep{AMICO2017201}, and we can visualize the
loadings by reshaping the vectors back to the symmetric matrix. The true
loadings of latent non-Gaussian components are plotted in figure
\ref{fig:origin}.

\begin{Schunk}
\begin{Sinput}
library(singR)
data(exampledata)
data <- exampledata

lgrid = 33
par(mfrow = c(2, 4))
# Components for X
image(matrix(data$sjX[1, ], lgrid, lgrid), col = heat.colors(12), xaxt = "n",
    yaxt = "n", main = expression("True S"["Jx"] * ", 1"))
image(matrix(data$sjX[2, ], lgrid, lgrid), col = heat.colors(12), xaxt = "n",
    yaxt = "n", main = expression("True S"["Jx"] * ", 2"))
image(matrix(data$siX[1, ], lgrid, lgrid), col = heat.colors(12), xaxt = "n",
    yaxt = "n", main = expression("True S"["Ix"] * ", 1"))
image(matrix(data$siX[2, ], lgrid, lgrid), col = heat.colors(12), xaxt = "n",
    yaxt = "n", main = expression("True S"["Ix"] * ", 2"))

# Components for Y
image(vec2net(data$sjY[1, ]), col = heat.colors(12), xaxt = "n", yaxt = "n",
    main = expression("True S"["Jy"] * ", 1"))
image(vec2net(data$sjY[2, ]), col = heat.colors(12), xaxt = "n", yaxt = "n",
    main = expression("True S"["Jy"] * ", 2"))
image(vec2net(data$siY[1, ]), col = heat.colors(12), xaxt = "n", yaxt = "n",
    main = expression("True S"["Iy"] * ", 1"))
image(vec2net(data$siY[2, ]), col = heat.colors(12), xaxt = "n", yaxt = "n",
    main = expression("True S"["Iy"] * ", 2"))
\end{Sinput}
\end{Schunk}

\begin{Schunk}
\begin{figure}
\includegraphics[width=1\linewidth]{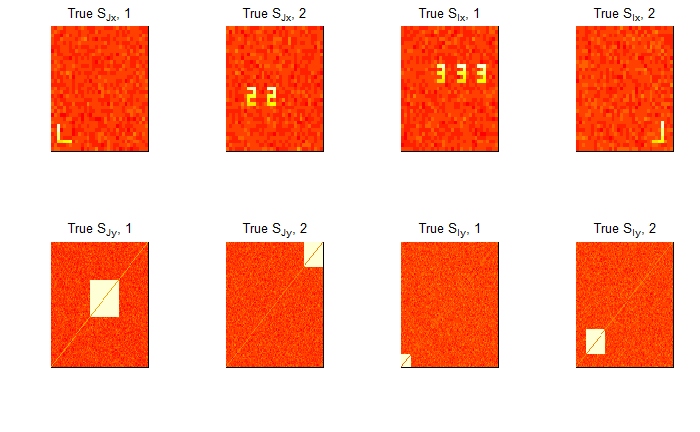} \caption[True loadings in example 1]{True loadings in example 1.}\label{fig:origin}
\end{figure}
\end{Schunk}

\textbf{Function singR performs all steps in the SING pipeline as a
single function}

We first illustrate the use of the wrapper function \texttt{singR} using
the default settings. We will describe optional arguments in more detail
in example 2.

\begin{Schunk}
\begin{Sinput}
example1 = singR(dX = data$dX, dY = data$dY, individual = T)
\end{Sinput}
\end{Schunk}

\textbf{Details of the SING pipeline}

We next explain each of the steps involved in SING estimation. Using
these individual functions in place of the high-level \texttt{singR}
function allows additional fine-tuning and can be helpful for large
datasets.

Estimate the number of non-Gaussian components in datasets dX and dY
using \texttt{FOBIasymp} from \CRANpkg{ICtest}:

\begin{Schunk}
\begin{Sinput}
n.comp.X = NG_number(data$dX)
n.comp.Y = NG_number(data$dY)
\end{Sinput}
\end{Schunk}

Apply \texttt{lngca} separately to each dataset using the JB statistic
as the measure of non-Gaussianity:

\begin{Schunk}
\begin{Sinput}
# JB on X
estX_JB = lngca(xData = data$dX, n.comp = n.comp.X, whiten = "sqrtprec",
    restarts.pbyd = 20, distribution = "JB")
Uxfull <- estX_JB$U
Mx_JB = est.M.ols(sData = estX_JB$S, xData = data$dX)

# JB on Y
estY_JB = lngca(xData = data$dY, n.comp = n.comp.Y, whiten = "sqrtprec",
    restarts.pbyd = 20, distribution = "JB")
Uyfull <- estY_JB$U
My_JB = est.M.ols(sData = estY_JB$S, xData = data$dY)
\end{Sinput}
\end{Schunk}

Use \texttt{greedymatch} to reorder \(\widehat{U}_{x}\) and
\(\widehat{U}_{y}\) by descending matched correlations and use
\texttt{permTestJointRank} to estimate the number of joint components:

\begin{Schunk}
\begin{Sinput}
matchMxMy = greedymatch(scale(Mx_JB, scale = F), scale(My_JB, scale = F),
    Ux = Uxfull, Uy = Uyfull)
permJoint <- permTestJointRank(matchMxMy$Mx, matchMxMy$My)
joint_rank = permJoint$rj
\end{Sinput}
\end{Schunk}

For preparing input to \texttt{curvilinear\_c}, manually prewhiten dX
and dY to get \(\widehat{L}_{x}^{-1}\) and \(\widehat{L}_{y}^{-1}\):

\begin{Schunk}
\begin{Sinput}
# Center X and Y
dX = data$dX
dY = data$dY
n = nrow(dX)
pX = ncol(dX)
pY = ncol(dY)
dXcentered <- dX - matrix(rowMeans(dX), n, pX, byrow = F)
dYcentered <- dY - matrix(rowMeans(dY), n, pY, byrow = F)

# For X Scale rowwise
est.sigmaXA = tcrossprod(dXcentered)/(pX - 1)
whitenerXA = est.sigmaXA %^% (-0.5)
xDataA = whitenerXA %*% dXcentered
invLx = est.sigmaXA %^% (0.5)

# For Y Scale rowwise
est.sigmaYA = tcrossprod(dYcentered)/(pY - 1)
whitenerYA = est.sigmaYA %^% (-0.5)
yDataA = whitenerYA %*% dYcentered
invLy = est.sigmaYA %^% (0.5)
\end{Sinput}
\end{Schunk}

Obtain a reasonable value for the penalty \(\rho\) by calculating the JB
statistics for all the joint components:

\begin{Schunk}
\begin{Sinput}
# Calculate the Sx and Sy.
Sx = matchMxMy$Ux[1:joint_rank, ] %*% xDataA
Sy = matchMxMy$Uy[1:joint_rank, ] %*% yDataA

JBall = calculateJB(Sx) + calculateJB(Sy)

# Penalty used in curvilinear algorithm:
rho = JBall/10
\end{Sinput}
\end{Schunk}

Estimate \(\widehat{U}_{x}\) and \(\widehat{U}_{y}\) with
\texttt{curvilinear\_c}:

\begin{Schunk}
\begin{Sinput}
# alpha=0.8 corresponds to JB weighting of skewness and kurtosis (can
# customize to use different weighting):
alpha = 0.8
# tolerance:
tol = 1e-10

out <- curvilinear_c(invLx = invLx, invLy = invLy, xData = xDataA, yData = yDataA,
    Ux = matchMxMy$Ux, Uy = matchMxMy$Uy, rho = rho, tol = tol, alpha = alpha,
    maxiter = 1500, rj = joint_rank)
\end{Sinput}
\end{Schunk}

Obtain the final result:

\begin{Schunk}
\begin{Sinput}
# Estimate Sx and Sy and true S matrix
Sjx = out$Ux[1:joint_rank, ] %*% xDataA
Six = out$Ux[(joint_rank + 1):n.comp.X, ] %*% xDataA
Sjy = out$Uy[1:joint_rank, ] %*% yDataA
Siy = out$Uy[(joint_rank + 1):n.comp.Y, ] %*% yDataA

# Estimate Mj and true Mj
Mxjoint = tcrossprod(invLx, out$Ux[1:joint_rank, ])
Mxindiv = tcrossprod(invLx, out$Ux[(joint_rank + 1):n.comp.X, ])
Myjoint = tcrossprod(invLy, out$Uy[1:joint_rank, ])
Myindiv = tcrossprod(invLy, out$Uy[(joint_rank + 1):n.comp.Y, ])

# signchange to keep all the S and M skewness positive
Sjx_sign = signchange(Sjx, Mxjoint)
Sjy_sign = signchange(Sjy, Myjoint)
Six_sign = signchange(Six, Mxindiv)
Siy_sign = signchange(Siy, Myindiv)

Sjx = Sjx_sign$S
Sjy = Sjy_sign$S
Six = Six_sign$S
Siy = Siy_sign$S

Mxjoint = Sjx_sign$M
Myjoint = Sjy_sign$M
Mxindiv = Six_sign$M
Myindiv = Siy_sign$M

est.Mj = aveM(Mxjoint, Myjoint)

trueMj <- data.frame(mj1 = data$mj[, 1], mj2 = data$mj[, 2], number = 1:48)
SINGMj <- data.frame(mj1 = est.Mj[, 1], mj2 = est.Mj[, 2], number = 1:48)
\end{Sinput}
\end{Schunk}

Plot \(\widehat{S}_{Jx}\), \(\widehat{S}_{Jy}\), \(\widehat{S}_{Ix}\),
and \(\widehat{S}_{Iy}\) in figure \ref{fig:estiexample1}.

\begin{Schunk}
\begin{figure}
\includegraphics[width=1\linewidth]{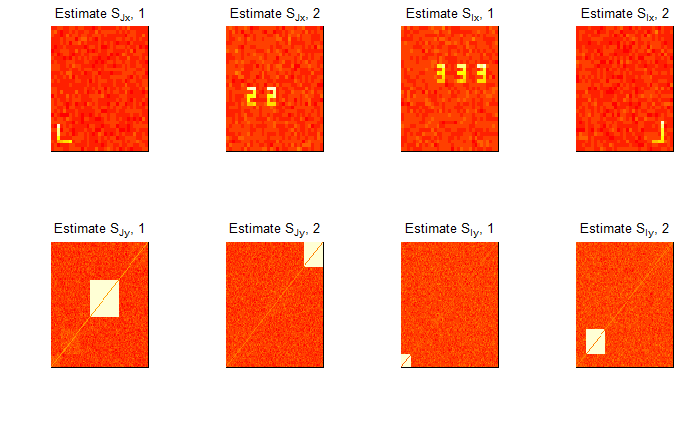} \caption[Estimated joint loadings in example 1]{Estimated joint loadings in example 1.}\label{fig:estiexample1}
\end{figure}
\end{Schunk}

Plot \(\widehat{M}_J\) in figure \ref{fig:mjex1}.

\begin{Schunk}
\begin{Sinput}
library(tidyverse)
library(ggpubr)

t1 <- ggplot(data = trueMj) + geom_point(mapping = aes(y = mj1, x = number)) +
    ggtitle(expression("True M"["J"] * ", 1")) + theme_bw() + theme(panel.grid = element_blank())

t2 <- ggplot(data = trueMj) + geom_point(mapping = aes(y = mj2, x = number)) +
    ggtitle(expression("True M"["J"] * ", 2")) + theme_bw() + theme(panel.grid = element_blank())

# SING mj

S1 <- ggplot(data = SINGMj) + geom_point(mapping = aes(y = mj1, x = number)) +
    ggtitle(expression("Estimated M"["J"] * ", 1")) + theme_bw() + theme(panel.grid = element_blank())

S2 <- ggplot(data = SINGMj) + geom_point(mapping = aes(y = mj2, x = number)) +
    ggtitle(expression("Estimated M"["J"] * ", 2")) + theme_bw() + theme(panel.grid = element_blank())

ggarrange(t1, t2, S1, S2, ncol = 2, nrow = 2)
\end{Sinput}
\end{Schunk}

\begin{Schunk}
\begin{figure}
\includegraphics[width=1\linewidth]{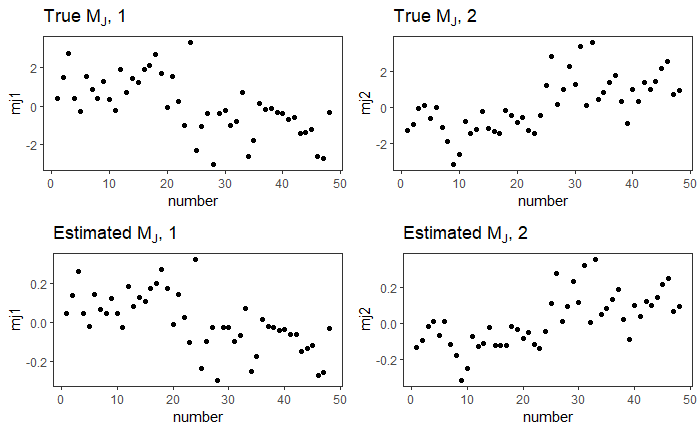} \caption[Estimated joint subject scores in example 1]{Estimated joint subject scores in example 1.}\label{fig:mjex1}
\end{figure}
\end{Schunk}

\hypertarget{example-2.-mri-data-simulation}{%
\subsection{Example 2. MRI data
simulation}\label{example-2.-mri-data-simulation}}

This example is a simulation inspired by the real data analysis of the
Human Connectome Project from \citet{RN146}. \(X\) are generated from
\(\widehat{S}_X\) from working memory task maps and \(Y\) are generated
from \(\widehat{S}_Y\) from resting-state correlations from a previous
SING analysis of the Human Connectome Project. The working memory
loadings are defined on the cortical surface, which is the highly folded
ribbon of gray matter forming the outer layer of the brain containing
billions of neurons. Large working memory loadings indicate locations in
the brain that tend to work together during memory tasks. The
resting-state correlation loadings are defined using a brain
parcellation from \citep{glasser2016multi} and
\citep{akiki2019determining}. Large resting-state loadings are related
to large correlations between brain regions occurring when a participant
is lying in a scanner performing no task. Additional details are in
\citep{RN146}. For the purposes of this example, we lower the resolution
of the working memory loadings and subset parts of the rs-correlation
loadings to reduce computation time. The \texttt{simdata} can be found
in the github repository \citep{singR_2022}.

\begin{Schunk}
\begin{Sinput}
# Load the package
library(singR)

# Read and visualize data
load("extdata/simdata.rda")

# sign change makes the skewness positive, which makes the region of
# 'activation' yellow in the plots that follow
Sxtrue = signchange(simdata$sjx)$S
Sytrue = signchange(simdata$sjy)$S
\end{Sinput}
\end{Schunk}

The \texttt{simdata.rda} have already been resampled from 32k to 2k
resolution to reduce computation time. Next, we resample the background
surface (i.e., template) to the same resolution, which will allow us to
plot the loadings on the cortical surface. This step uses
\CRANpkg{ciftiTools} \citep{pham2021ciftitools} and connectome workbench
\citep{marcus2011informatics}. To run this code, one needs to install
connectome workbench, as described in
(\url{https://github.com/mandymejia/ciftiTools}).

\begin{Schunk}
\begin{Sinput}
library(ciftiTools)
ciftiTools.setOption("wb_path", "C:/Software/workbench")

xii_template <- read_cifti("extdata/template.dtseries.nii", brainstructures = c("left",
    "right"), resamp_res = 2000)
## the template cifti file is built in the package, and resample to
## 2k resolution.

xii_new <- newdata_xifti(xii_template, t(Sxtrue))
view_xifti_surface(select_xifti(xii_new, 1), zlim = c(-2.43, 2.82))  ## true S_JX1
view_xifti_surface(select_xifti(xii_new, 2), zlim = c(-2.43, 2.82))  ## true S_JX2
\end{Sinput}
\end{Schunk}

\begin{Schunk}
\begin{figure}
\includegraphics[width=1\linewidth]{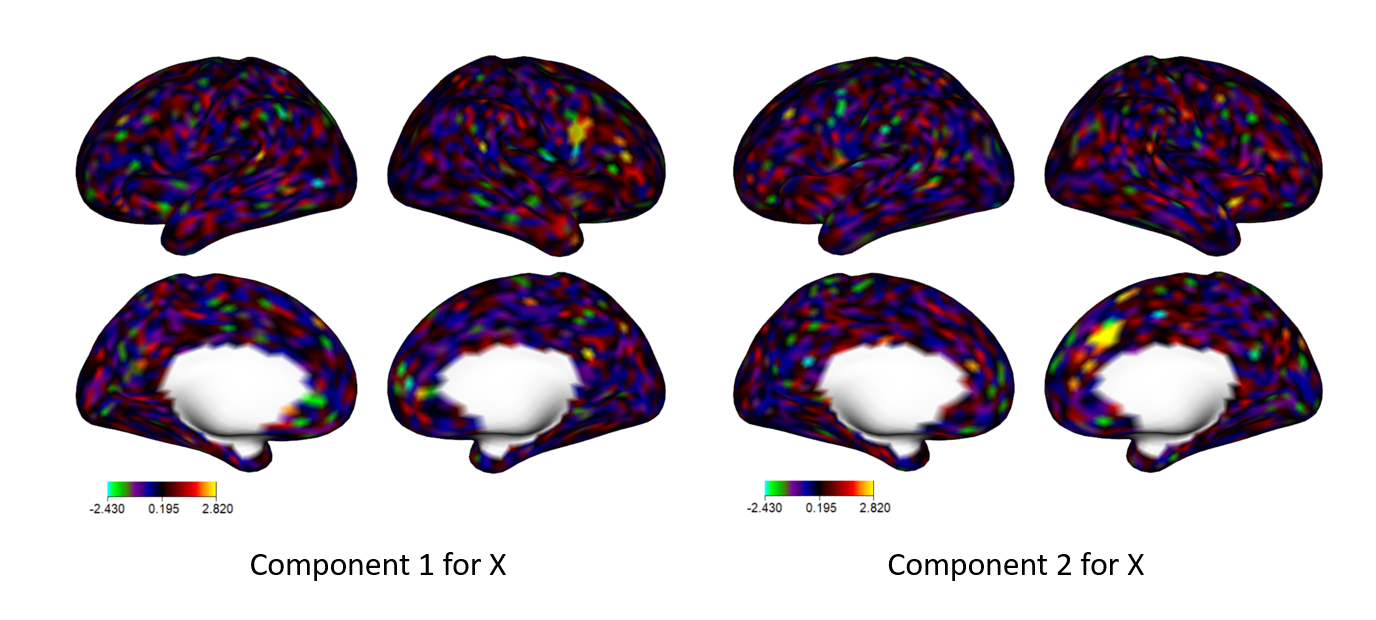} \caption[True joint loadings in SJX]{True joint loadings in SJX}\label{fig:truex}
\end{figure}
\end{Schunk}

In figure \ref{fig:truex}, the yellow regions indicate locations with
large loadings. Similar plots can be created for the two individual
components (not shown). When applied to fMRI activation maps, SING tends
to identify small patches of cortex, similar to this figure.

Next, we convert the rows of \(S_{Jy}\) to symmetric matrices and create
plots. The nodes are organized into communities (i.e., modules) to aid
visualization. SING tends to identify a single node and the connections
with this node, which result in a cross-like pattern in the matrix
representation. The joint loadings are plotted in figure \ref{fig:truey}
with \texttt{plotNetwork\_change}, which is defined below. Similar plots
can be created for the loadings from the two individual components.

\begin{Schunk}
\begin{Sinput}
# define plotNetwork_change
plotNetwork_change = function(component, title = "", qmin = 0.005, qmax = 0.995,
    path = "mmpplus.csv", make.diag = NA) {
    # component: vectorized network of length choose(n,2)
    require(ggplot2)
    require(grid)
    require(scales)

    # load communities for plotting:
    mmp_modules = read.csv(path, header = TRUE)
    mmp_order = order(mmp_modules$Community_Vector)

    zmin = quantile(component, qmin)
    zmax = quantile(component, qmax)

    netmat = vec2net(component, make.diag)

    meltsub = create.graph.long(netmat, mmp_order)

    g2 = ggplot(meltsub, aes(X1, X2, fill = value)) + geom_tile() + scale_fill_gradient2(low = "blue",
        high = "red", limits = c(zmin, zmax), oob = squish) + labs(title = title,
        x = "Node 1", y = "Node 2") + coord_cartesian(clip = "off", xlim = c(-0,
        100))

    loadingsummary = apply(abs(netmat), 1, sum, na.rm = TRUE)
    loadingsum2 = loadingsummary[mmp_order]

    Community = factor(mmp_modules$Community_Label)[mmp_order]

    g3 = qplot(c(1:100), loadingsum2, col = Community, size = I(3)) + xlab("MMP Index") +
        ylab("L1 Norm of the Rows")

    return(list(netmatfig = g2, loadingsfig = g3, netmat = netmat, loadingsummary = loadingsummary))
}
\end{Sinput}
\end{Schunk}

\begin{Schunk}
\begin{Sinput}
library(cowplot)
# plot for the true component of Y
path = "extdata/new_mmp.csv"
out_true1 = plotNetwork_change(Sytrue[1, ], title = expression("True S"["Jx"] *
    ", 1"), qmin = 0.005, qmax = 0.995, path = path)
out_true2 = plotNetwork_change(Sytrue[2, ], title = expression("True S"["Jx"] *
    ", 2"), qmin = 0.005, qmax = 0.995, path = path)

p1 = out_true1$netmatfig
p2 = out_true1$loadingsfig
p3 = out_true2$netmatfig
p4 = out_true2$loadingsfig

plot_grid(p1, p2, p3, p4, nrow = 2)
\end{Sinput}
\end{Schunk}

\begin{Schunk}
\begin{figure}
\includegraphics[width=1\linewidth]{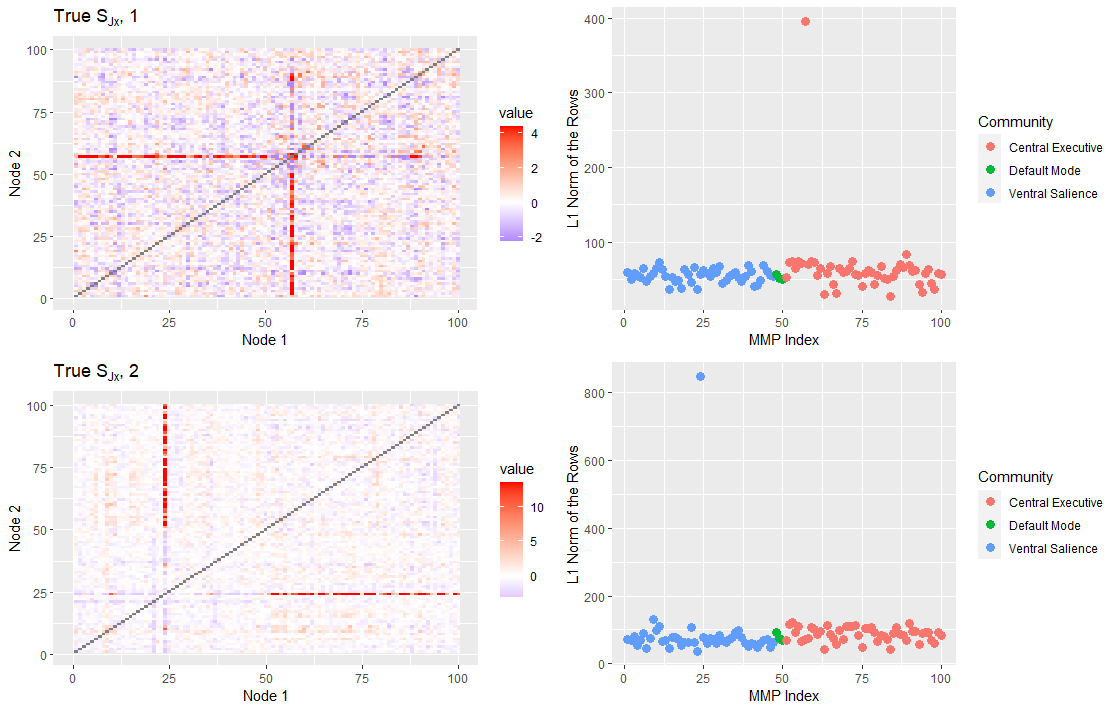} \caption[True joint loadings in SJY in example 2]{True joint loadings in SJY in example 2}\label{fig:truey}
\end{figure}
\end{Schunk}

\textbf{Function singR performs all steps in the SING pipeline as a
single function}

In example 1, we introduced the pipeline of the SING method. We will use
example 2 to explain the \texttt{singR} function in detail. The default
output of \texttt{singR} is a list of \(\widehat{S}_{Jx}\),
\(\widehat{S}_{Jy}\), \(\widehat{M}_{J}\), \(\widehat{M}_{Jx}\),
\(\widehat{M}_{Jy}\). By default, it will center the data such that the
mean of each row is equal to zero. In our simulated dataset, all
variables are on the same scale, and consequently we do not perform
standardization (\texttt{stand=FALSE}). When \texttt{stand=TRUE}, the
data are additionally standardized to have the mean and variance of each
column equal to zero, which is the standardization commonly used in PCA.
If \texttt{n.comp.X} and \texttt{n.comp.Y} are not specified,
\texttt{singR} will use \texttt{FOBIasymp} from \CRANpkg{ICtest} to
estimate the number of non-Gaussian components in each dataset, which
requires additional computational expense. (Other tests of the number of
non-Gaussian components accounting for spatial smoothing/autocorrelation
can be found in \citep{zhao2022group}, which may be more effective for
spatially correlated data but are generally slower.)

When \texttt{individual\ =\ TRUE}, the \texttt{singR} will additionally
output \(\widehat{M}_{Ix}\), \(\widehat{M}_{Iy}\), \(\widehat{S}_{Ix}\),
and \(\widehat{S}_{Iy}\). When
\texttt{distribution\ =\ "tiltedgaussian"}, non-Gaussian components will
be estimated through \texttt{lngca} using likelihood component analysis,
which is slower but can be more accurate. By default
\texttt{distribution\ =\ "JB"}, and \texttt{lngca} will use the
Jarque-Bera test statistic as the measure of non-Gaussianity of each
component.

The \texttt{Cplus} argument determines whether to use
\texttt{curvilinear\_c} or \texttt{curvilinear} in \texttt{singR}.
\texttt{curvilinear} is implemented with pure R but is slow while
\texttt{curvilinear\_c} uses C++. The parameter \texttt{rho\_extent} can
be one of \texttt{c("small",\ "medium"\ or\ "large")} or a number. This
determines the penalty \(\rho\) in \texttt{curvilinear} or
\texttt{curvilinear\_c} that results in equal or highly correlated
\(\widehat{M}_{Jx}\) and \(\widehat{M}_{Jy}\). Additionally, we can use
\texttt{pmse()} to evaluate the distance between two subject scores
matrices. With larger \(\rho\), the
\(pmse\left(\widehat{M}_{Jx}, \widehat{M}_{Jy}\right)\) value will be
smaller. Usually, ``small'' \(\rho\) is sufficient for approximately
equal \(\widehat{M}_{Jx}\) and \(\widehat{M}_{Jy}\). We have observed
that very large \(\rho\) can adversely impact the accuracy of the
loadings. Our recommendation is to use ``small'' \(\rho\) and check if
it results in equal scores, and if not, then try other settings. The
code below took approximately 20 seconds to run on a 2.8 GHz processor,
but may take longer depending on the processing speed.

\begin{Schunk}
\begin{Sinput}
example2 = singR(dX = simdata$dX, dY = simdata$dY, rho_extent = "small",
    Cplus = TRUE, stand = FALSE, individual = TRUE, distribution = "JB")
\end{Sinput}
\end{Schunk}

The joint loadings \(\widehat{S}_{JX}\) are depicted in figure
\ref{fig:estx2}.

\begin{Schunk}
\begin{Sinput}
xii_new <- newdata_xifti(xii_template, t(example2$Sjx))

view_xifti_surface(select_xifti(xii_new, 1), zlim = c(-2.43, 2.82))  ## component1 small rho
view_xifti_surface(select_xifti(xii_new, 2), zlim = c(-2.43, 2.82))  ## component2 small rho
\end{Sinput}
\end{Schunk}

\begin{Schunk}
\begin{figure}
\includegraphics[width=1\linewidth]{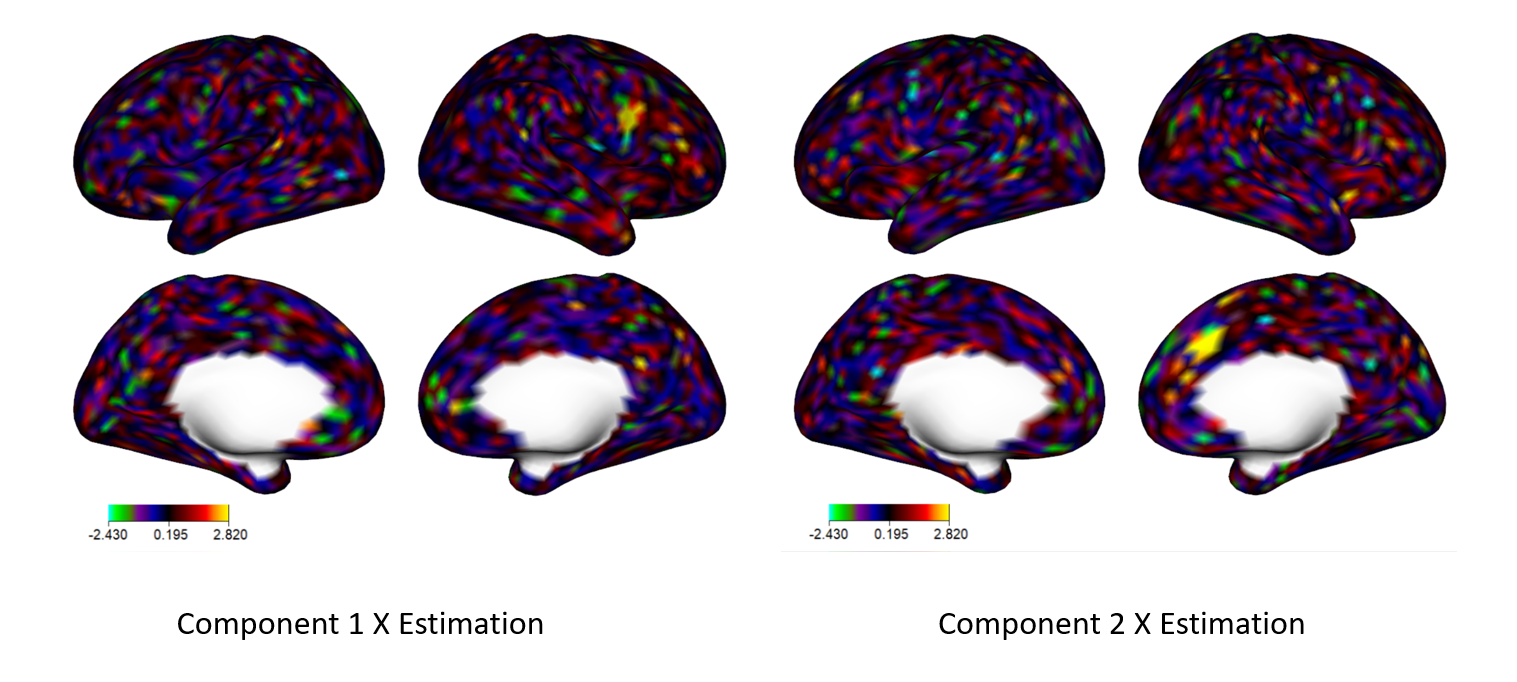} \caption[Estimated joint loadings in SJX in example 2]{Estimated joint loadings in SJX in example 2.}\label{fig:estx2}
\end{figure}
\end{Schunk}

The joint loadings \(\widehat{S}_{Jy}\) are depicted in figure
\ref{fig:esty2}.

\begin{Schunk}
\begin{Sinput}
library(cowplot)
path = "extdata/new_mmp.csv"
out_rhoSmall1 = plotNetwork_change(example2$Sjy[1, ], title = expression("Estimate S"["Jy"] *
    ", 1"), qmin = 0.005, qmax = 0.995, path = path)
out_rhoSmall2 = plotNetwork_change(example2$Sjy[2, ], title = expression("Estimate S"["Jy"] *
    ", 2"), qmin = 0.005, qmax = 0.995, path = path)

p5 = out_rhoSmall1$netmatfig
p6 = out_rhoSmall1$loadingsfig
p7 = out_rhoSmall2$netmatfig
p8 = out_rhoSmall2$loadingsfig

plot_grid(p5, p6, p7, p8, nrow = 2)
\end{Sinput}
\end{Schunk}

\begin{Schunk}
\begin{figure}
\includegraphics[width=1\linewidth]{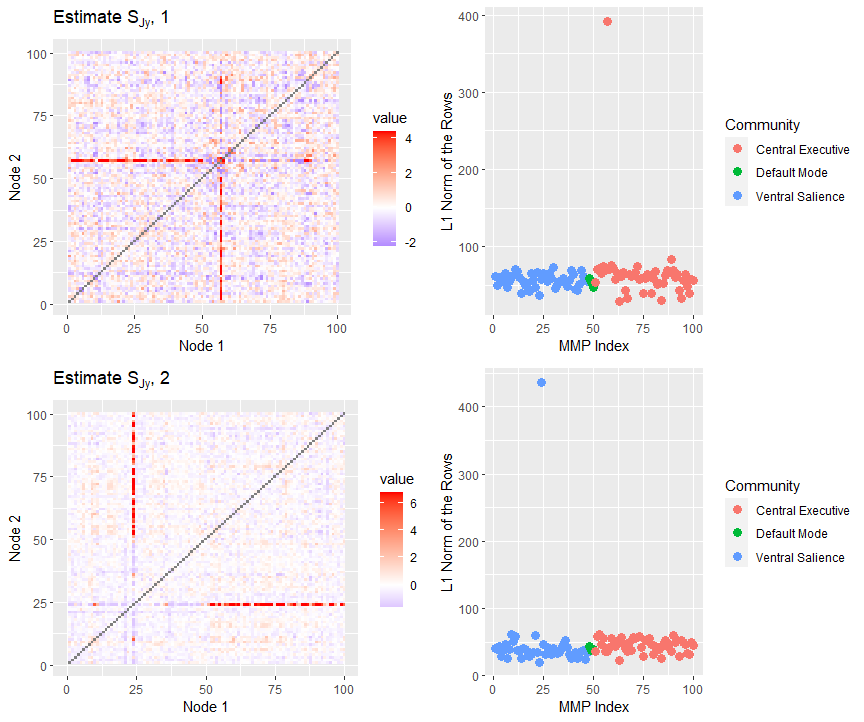} \caption[Estimated joint loadings in SJY in example 2]{Estimated joint loadings in SJY in example 2.}\label{fig:esty2}
\end{figure}
\end{Schunk}

\hypertarget{summary}{%
\section{Summary}\label{summary}}

This paper introduces the \CRANpkg{singR} package and demonstrates how
simultaneous non-Gaussian component analysis can be used to extract
shared features from two datasets using R. The main contribution of the
R package \CRANpkg{singR} is to provide easy code for data integration
in neuroscience. We introduce the function \texttt{singR}, which
combines the SING pipeline into one function that performs data
standardization, estimates the number of non-Gaussian components and
estimates the number of joint components. Previous analyses indicate the
joint structure estimated by SING can improve upon other neuroimaging
data integration methods. SING can reveal new insights by using
non-Gaussianity for both dimension reduction and latent variable
extraction, whereas ICA methods involve an initial PCA step that tends
to aggregate features and can remove information.

\hypertarget{acknowledgments}{%
\section{Acknowledgments}\label{acknowledgments}}

Research reported in this publication was supported by the National
Institute of Mental Health of the National Institutes of Health under
award number R01MH129855 to BBR. The research was also supported by the
Division of Mathematical Sciences of the National Science Foundation
under award number DMS-2044823 to IG. The content is solely the
responsibility of the authors and does not necessarily represent the
official views of the National Institutes of Health and National Science
Foundation.

Simulated data were based on a previous analysis of data from the Human
Connectome Project. These data were provided {[}in part{]} by the Human
Connectome Project, WU-Minn Consortium (Principal Investigators: David
Van Essen and Kamil Ugurbil; 1U54MH091657) funded by the 16 NIH
Institutes and Centers that support the NIH Blueprint for Neuroscience
Research; and by the McDonnell Center for Systems Neuroscience at
Washington University.

\bibliography{filename}

\address{%
Liangkang Wang\\
Emory University\\%
Department of Biostatistics and Bioinformatics\\ Atlanta, Georgia, US\\
\textit{ORCiD: \href{https://orcid.org/0000-0003-3393-243X}{0000-0003-3393-243X}}\\%
\href{mailto:liangkang.wang@emory.edu}{\nolinkurl{liangkang.wang@emory.edu}}%
}

\address{%
Irina Gaynanova\\
Texas A\&M University\\%
Department of Statistics\\ College Station, Texas, US\\
\url{https://irinagain.github.io/}\\%
\textit{ORCiD: \href{https://orcid.org/0000-0002-4116-0268}{0000-0002-4116-0268}}\\%
\href{mailto:irinag@tamu.edu}{\nolinkurl{irinag@tamu.edu}}%
}

\address{%
Benjamin Risk\\
Emory University\\%
Department of Biostatistics and Bioinformatics\\ Atlanta, Georgia, US\\
\url{https://github.com/thebrisklab/}\\%
\textit{ORCiD: \href{https://orcid.org/0000-0003-1090-0777}{0000-0003-1090-0777}}\\%
\href{mailto:benjamin.risk@emory.edu}{\nolinkurl{benjamin.risk@emory.edu}}%
}

\end{article}

\end{document}